\documentclass[12pt,a4paper]{article}

\usepackage{epsfig} 
\usepackage{graphicx}

\begin{document}
\textwidth=135mm
 \textheight=200mm

\begin{center}
{\bfseries Multimessenger search for evaporating primordial black holes
\footnote{{\small Talk at the International Conference ``SN 1987A, Quark Phase Transition
in Compact Objects and Multimessenger Astronomy'', KBR, Terskol (BNO) -- KChR, Nizhnij Arkhyz (SAO), Russia,
July 2 -- 8, 2017}}
}

\vskip 5mm

V.~B.~Petkov$^{1, 2}$, E.~V.~Bugaev$^1$, P.~A.~Klimai$^1$   
\vskip 5mm

{\small {\it $^1$ Institute for Nuclear Research of RAS, Moscow, Russia}} \\
{\small {\it $^2$ Institute of Astronomy of  RAS, Moscow, Russia}}
\\
\end{center}
\vskip 5mm

\centerline{\bf Abstract}
Primordial black holes (PBHs) are black holes which may form in the early Universe through the gravitational collapse of primordial cosmological density fluctuations. Due to Hawking radiation these PBHs are supposed to evaporate by emitting particles. Recent developments in the experimental searching for evaporating PBHs in the local Universe are reviewed. The multimessenger techniques of searching for signals from evaporating PBHs are discussed. 

\vskip 10mm
\section{\label{sec:intro}Introduction}
Primordial black holes can be formed in the early Universe through the gravitational collapse of primordial cosmological density fluctuations -- those that give rise to the observed structure of the Universe (galaxies and clusters of galaxies) during its subsequent evolution. For an appreciable number of PBHs to be formed, it is important that significant density fluctuations on small mass scales existed in the early Universe. The curvature fluctuations and the related density fluctuations are currently believed to result from an inflationary expansion of the Universe; significantly, the power spectrum of these fluctuations is entirely determined by the parameters of the theoretical inflation model used and primarily by the form of the inflation potential. There exist quite a few models (see, e.g., \cite{1_Bugaev08} and references therein) in which a fluctuation spectrum that ensures the formation of a considerable number of PBHs is predicted.
 
The regularities of the black hole formation are determined not only by the cosmology and physics of the early Universe. Theoretical predictions of the PBH formation probability depend strongly on the theory of gravitation and the model of gravitational collapse used. Direct search for the PBHs is based on the Hawking radiation \cite{2_Hawking74, 3_Hawking75}, which leads to their evaporation on a characteristic time scale \cite{4_Carr16, 5_Ukwatta16} 
\begin{equation}
\tau(M_{BH}) \simeq 1.15 \cdot 10^{65} \left(\frac{M_{BH}}{M_{Sun}}\right)^{3} y. 
\label{Eq1}\end{equation}

The critical mass for which $\tau$ equals the age of the Universe ($\simeq$ 13.7 Gyr) is $\simeq 5.1 \cdot 10^{14}$ g \cite{4_Carr16}.
It should be noted that the evaporation of black holes has not been completely studied to date. There are several theoretical models of the evaporation process \cite{6_MacGibbon90} --  \cite{10_Lalakulich04}. The technique of searching for the high energy photon signal from evaporating PBHs depends on temporal and energy characteristics of their gamma-ray emission. Because these characteristics differ for different evaporation models, the upper limit obtained for the number density of evaporating PBHs in a local region of space depends strongly on the specific evaporation model.
Of course the distribution of PBHs in space is important for their direct search. Because of the local increase in the density of PBHs in our Galaxy \cite{11_Chisholm06}, the constraints on their number density imposed by a direct search can be more stringent than those imposed by diffuse extragalactic gamma-ray background measurements, which are sensitive only to the mean PBH density in the Universe.

PBHs might arguably be the most natural candidates to solve the dark matter problem: they are cold, weakly-interacting, and do not require extensions of the Standard Model of particle physics. So, experimental detection of PBHs could provide a unique probe of the early Universe, gravitational collapse, high energy physics and quantum gravity. The no detection of PBHs at the current level of the experimental technique also carries useful information and allows further progress to be made in understanding the early Universe.

\section{Search for gamma-ray bursts from evaporating PBHs}
At the final PBH evaporation stage the high-energy gamma-ray burst (significant and time-localized excess of gamma radiation above the background) is generated. Different evaporation models give different temporal and energy characteristics of these bursts. Experimental search for such events has been carried out at several EAS arrays and Cherenkov telescopes for three evaporation models. In the first (best-known) model \cite{6_MacGibbon90}, the photons produced by the fragmentation of evaporated quarks are assumed to make a large contribution to the total photon spectrum from evaporating PBHs. In the other two models \cite{7_Heckler97, 8_Daghigh02}, the photons produced by the interaction of evaporated quarks (and leptons) with one another are also taken into account. The interactions of evaporated particles are important if something like a photo- or chromosphere is formed around the PBH during its evaporation (as is assumed in  \cite{7_Heckler97, 8_Daghigh02}). In these experiments the upper limits on the number density of evaporating PBHs in local region of the Galaxy have been obtained \cite{12_Petkov15}.

According to model with quark gluon phase transition \cite{9_Cline97}, the gamma-ray burst occurs when in the vicinity of a black hole in the flow of its radiation the hadron-quark phase transition take place, which can happen at $T~>~100$ MeV. The quark-gluon plasma, once created, absorbs the radiation of ever increasing temperature, emitted by the black hole. Then the energy accumulated in the shell of plasma is ejected for a short period of time ($\sim$ 100 ms) as a burst of $\sim$ 100 keV photons. Their mechanism could, in principle, explain some experimentally detected rather short gamma-ray bursts \cite{9_Cline97}.

The evaporation model with relativistic phase transitions predicts ultrashort ($\sim 10^{-13}$ s) gamma-ray bursts with the spectrum with the maximum intensities simultaneously at the photon energies of 100 MeV and 100 GeV \cite{10_Lalakulich04}. Such ultrashort gamma-ray bursts can be detected by EAS arrays located on mountains as EASs with a uniform lateral distribution. Experimental search for PBHs in frame of this evaporation model was carried out at Andyrchy EAS array; a limit on the concentration of evaporating PBHs in a local region of the Galaxy for this evaporation model has been obtained \cite{14_Vereshkov15}. 

So far the searching for very-high-energy gamma-ray bursts from evaporating PBH yields only limits on their concentration in local region of the Galaxy. Moreover, in this kind of experiments only limits on the PBH's concentration can be obtained. In order to prove that particular event is the burst from evaporating PBH the multimessenger approach is needed.

\section{Multimessenger search for PBH evaporation signal with AMON \cite{15_Tesic15}}
Most evaporation models predict, after stage of gradual evaporation, an explosion of PBHs during the last few seconds of their lives. A burst of high energy particles is produced as a result of the explosion. Different kinds of these particles could be registered in coincidence by several detectors with large fields of view. The search for such events has been proposed within the Astrophysical Multimessenger Observatory Network (AMON) framework \cite{15_Tesic15}. The discovery potential for joint detections of multimessenger bursts due to evaporating PBHs has examined for IceCube (neutrinos), HAWC (gamma rays) and Pierre Auger (gamma rays, neutrons and protons). Only model without a chromosphere \cite{6_MacGibbon90} ] was considered for such search, because a PBH chromosphere  \cite{7_Heckler97, 8_Daghigh02} would give steeper particle spectra making these high-energy experiments not suitable for detection of the PBH bursts \cite{16_Petkov10}. It was argued that this multimessenger approach is essential to distinguish between bursts due to PBHs and other possible sources, should a positive detection occur. Both real-time and archival searches for PBH bursts are planned within the AMON framework.

\section{Transient radio and optical pulses from exploding PBHs}

\begin{figure}[tph!]
\centerline{\psfig{file=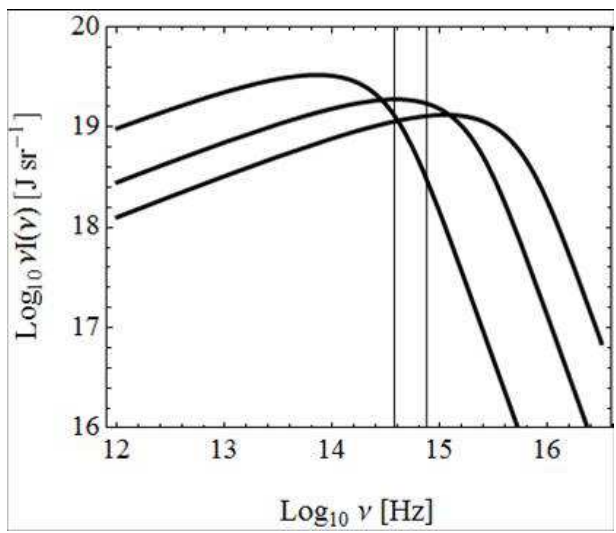,width=8cm}}
    \caption{Radiation spectra of exploding PBH calculated for Rees - Blandford mechanism \cite{17_Rees77, 18_Blanford77} (see text for detais). From left to right, Lorentz factor $\gamma= (4, 7, 10)\times 10^6$. Vertical lines bound the range of wavelength 400 -- 800 nm.}
\end{figure}

The electromagnetic pulses can be generated during PBHs explosions due to interactions of emitted charged particles (mainly electrons and positrons) with the interstellar magnetic field. Possible radiation mechanism, proposed by Rees \cite{17_Rees77}, consists in collective interaction of electrons and positrons with an ambient field. The emitted charged particles in this case are considered as a conducting sphere, expanding into a uniform magnetic field. The spectrum of radiation depends on Lorentz factor of expanding shell and strength of ambient magnetic field \cite{18_Blanford77}. Calculated spectra of electromagnetic emission for field strength of 0.5 nT are presented on Fig. 1 for three values of Lorentz factor. One can see that this radiation mechanism gives us possibility to detect exploding PBHs by the use of optical and/or radio telescopes (or both, in coincidence).

The search for radio pulses from PBH explosions was carried out using the Eight-meter-wavelength Transient Array (ETA) \cite{19_Cutchin15}. No compelling astrophysical signal was detected in this experiment; only upper limit on the rate of exploding PBHs was obtained for an exploding PBH with a fireball Lorentz factor of $10^{4.3}$.
 
It should be noted that the evaporation process of black holes is essentially changed in the presence of an extra spatial dimension. The fact of the matter is that with the addition of an extra spatial dimension, black holes could exist in different phases and undergo phase transitions \cite{20_Kol02, 21_Kavic08}. For one toroidally compactified extra dimension, two possible phases are a black string wrapping the compactified extra dimension, and a 5-dimensional black hole smaller than the extra dimension. A topological phase transition from the black string to the black hole results in a significant release of energy by means of Rees-Blandford mechanism. The ETA observations  \cite{19_Cutchin15} also imply an upper limit on the rate of PBH explosions in the context of certain extra dimension models as described in \cite{21_Kavic08}.

\section{Search for very high-energy gamma-ray bursts from evaporating PBHs in coincidence with optical flashes}
\begin{figure}[tph!]
\centerline {\psfig{file=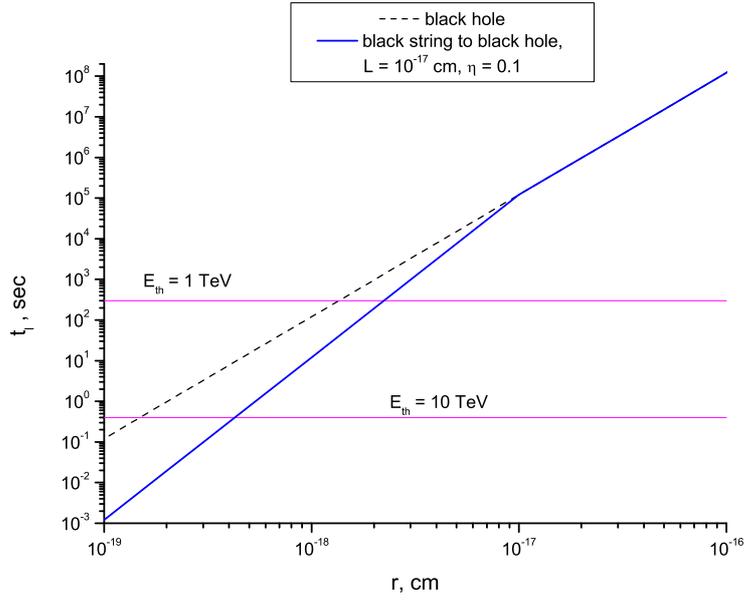,width=12cm}} 
\caption{The dependence of the time until the end of PBH evaporation tl on black hole radius r. Dashed line -- for usual space, solid line -- for space with one toroidally compactified extra dimension with $L = 10^{-19}$ m, where a topological phase transition from the black string to the black hole is happened. Horizontal lines show values of $t_l$ for two threshold energies of gamma-rays, 1 TeV and 10 TeV, for the model without a chromosphere.}
\end{figure}

As it was mentioned above, the multimessenger approach gives us a possibility to distinguish between bursts due to PBHs and other possible sources. The joint search for very high-energy gamma-ray bursts and optical flashes from evaporating PBHs is a kind of multimessenger approach. At present the quick search for astrophysical objects which produce both, bursts of high energy cosmic radiation and optical flashes, is carried out in the near real-time mode with the facilities of the Baksan Neutrino Observatory (BNO) of INR RAS and a complex of astronomical telescopes at the Terskol Peak Observatory (Terskol branch of INASAN) \cite{22_Kurenya18}. Unique complex of BNO facilities consists of Baksan Underground Scintillation Telescope (BUST) \cite{23_Alekseyev79, 24_Alekseyev98} and two EAS arrays: ``Carpet-2'' \cite{25_Dzhappuev07} and ``Andyrchy'' \cite{26_Petkov06}. The BNO facilities work in continuous mode of operation and they are recording of cosmic rays from upper hemisphere (so called ``all sky all time'' mode). These apparatus allow searching for bursts of cosmic gamma radiation in wide range of primary gamma-rays energy: from 1 TeV (at the BUST) up to 80 TeV (at the EAS arrays ``Carpet-2'' and ``Andyrchy'') \cite{27_Smirnov05, 28_Smirnov06, 29_Petkov08}. 

The search for very high-energy gamma-ray bursts from evaporating PBHs can be performed at EAS arrays only in frame of model without a chromosphere, because a PBH chromosphere would give steeper particle spectra making these high-energy experiments not suitable for detection of the PBH bursts \cite{16_Petkov10}. But even in this model the burst duration is very short ($\le 40$ ms) for the EAS arrays ``Carpet-2'' and ``Andyrchy'', due to their high energy thresholds \cite{29_Petkov08}. Therefore the burst of very high-energy gamma-rays and optical flash from the final black hole explosion happens in usual space practically simultaneously. However the burst of gamma-rays can be registered prior to optical flash at the BUST, with its lower threshold energy (see Fig. 2).
 
In the model with one toroidally compactified extra dimension, after topological phase transition from the black string to the black hole, a 5d black hole continues to evaporate with a different rate (Fig. 2). In principle two optical signals could be expected in this model: first one due to a topological phase transition from the black string to the black hole (when $L \sim r$) and second one from the final 5d black hole explosion, with delay between two successive signals depending on size of an extra dimension L. But second signal is expected to be weaker than that first one, with much smaller total emitted energy \cite{20_Kol02}. In any case ``all sky'' optical telescopes are needed, both for the successful search of the evaporating PBHs and for distinction of the evaporation models.

\vspace{5mm}

{\bf Acknowledgements.}\\
This study is performed with a part of the instrument certified as a Unique Scientific Facility (Baksan Underground Scintillation Telescope) and at an office that is an item of the Shared Research Facilities state program (Baksan Neutrino Observatory of the Institute for Nuclear Research). The work is supported by the Russian Foundation for Basic Research, project number 16-29-13034.

\newpage

\end{document}